\documentclass[10pt,journal,compsoc]{IEEEtran}

\ifCLASSOPTIONcompsoc
  \usepackage[nocompress]{cite}
\else
  \usepackage{cite}
\fi
\ifCLASSINFOpdf
\usepackage[pdftex]{graphicx}
	\graphicspath{{../img/}}

\else
  \usepackage[dvips]{graphicx}
  \graphicspath{{../img/}}
\fi
\usepackage{amsmath}
\usepackage{bbold}
\usepackage{algorithmic}
\usepackage{array}
\usepackage{url}
\usepackage{array,multirow}
\usepackage{algorithm}
\usepackage{algorithmic}
\usepackage{color}

\newcommand{\SWITCH}[1]{\STATE \textbf{switch} (#1)}
\newcommand{\ENDSWITCH}{\STATE \textbf{end switch}}
\newcommand{\CASE}[1]{\STATE \textbf{case} #1\textbf{:} \begin{ALC@g}}
\newcommand{\ENDCASE}{\end{ALC@g}}

\newcommand{\DEFAULT}{\STATE \textbf{default:} \begin{ALC@g}}
\newcommand{\ENDDEFAULT}{\end{ALC@g}}
\newcommand{\DEFAULTLINE}[1]{\STATE \textbf{default:} }

\ifCLASSOPTIONcompsoc
  \usepackage[caption=false,font=footnotesize,labelfont=sf,textfont=sf]{subfig}
\else
  \usepackage[caption=false,font=footnotesize]{subfig}
\fi

\begin{document}

\title{Non-Negative Matrix Factorizations for Multiplex Network Analysis}

\author{Vladimir~Gligorijevi\'c,
        Yannis~Panagakis,~~\IEEEmembership{Member,~IEEE}
        and Stefanos~Zafeiriou,~~\IEEEmembership{Member,~IEEE}% <-this % stops a space
\IEEEcompsocitemizethanks{\IEEEcompsocthanksitem V. Gligorijevi\'c, Y. Panagakis and S.  Zafeiriou are with the Department of Computing, Imperial College London, UK. Corresponding author: V. Gligorijevi\'c, email: v.gligorijevic@imperial.ac.uk}% <-this % stops an unwanted space
}


\markboth{Journal of \LaTeX\ Class Files,~Vol.~14, No.~8, August~2015}%
{Shell \MakeLowercase{\textit{et al.}}: Bare Demo of IEEEtran.cls for Computer Society Journals}

\IEEEtitleabstractindextext{%
\begin{abstract}

Networks have been a general tool for representing, analyzing, and modeling relational data arising in several domains. One of the most important aspect of network analysis is community detection or network clustering. 
Until recently, the major focus have been on discovering community structure in single (i.e., monoplex) networks. However, with the advent of relational data with multiple modalities, multiplex networks, i.e., networks composed of multiple layers representing different aspects of relations, have emerged. Consequently, community detection in multiplex network, i.e., detecting clusters of nodes shared by all layers, has become a new challenge. In this paper, we propose {\bf N}etwork {\bf F}usion for {\bf C}omposite {\bf C}ommunity {\bf E}xtraction ({\bf NF-CCE}), a new class of algorithms, based on four different non--negative matrix factorization models, capable of extracting composite communities in multiplex networks. Each algorithm works in two steps: first, it finds a non--negative, low--dimensional feature representation of each network layer; then, it fuses the feature representation of layers into a common non--negative, low--dimensional feature representation via collective factorization. The composite clusters are extracted from the common feature representation. We demonstrate the superior performance of our algorithms over the {\em state--of--the--art} methods on various types of multiplex networks, including biological, social, economic, citation, phone communication, and brain multiplex networks.
\end{abstract}

}

\maketitle

\IEEEdisplaynontitleabstractindextext

% \fi
\IEEEpeerreviewmaketitle

\IEEEraisesectionheading{\section{Introduction}\label{sec:introduction}}

\IEEEPARstart{N}{etworks} (or graphs\footnote{we use terms {\em graphs} and {\em network} interchangeably throughout this paper}) along with their theoretical foundations are powerful mathematical tools for representing, modeling, and analyzing complex systems arising in several scientific disciplines including sociology, biology, physics, and engineering among others \cite{Newman2010book}. Concretely, social networks, economic networks, biological networks, telecommunications networks, etc. are just a few examples of graphs in which a large set of entities (or agents) correspond to {\em nodes} (or {\em vertices}) and relationships or interactions between entities correspond to \textit{edges} (or {\em links}). Structural analysis of these networks have yielded important findings in the corresponding fields \cite{BoccalettiReview, Strogatz2001}.

{\em Community detection} (also known as {\em graph clustering} or {\em module detection}) is one of the foremost problems in network analysis. It aims to find groups of nodes (i.e., {\em clusters, modules} or {\em communities}) that are more densely connected to each other than they are to the rest of the network \cite{Fortunato2010}. Even thought the volume of research on community detection is large, e.g., \cite{Girvan2002, Pereira2004, Leskovec2010, Chen2006, Duch2005, Blondel2008, Mitrovic2009, Porter2009, Schaeffer2007}, the majority of these methods  focus on networks with only one type of relations between nodes (i.e., networks of single type interaction). 

\begin{figure}
\centering
\includegraphics[scale=0.3]{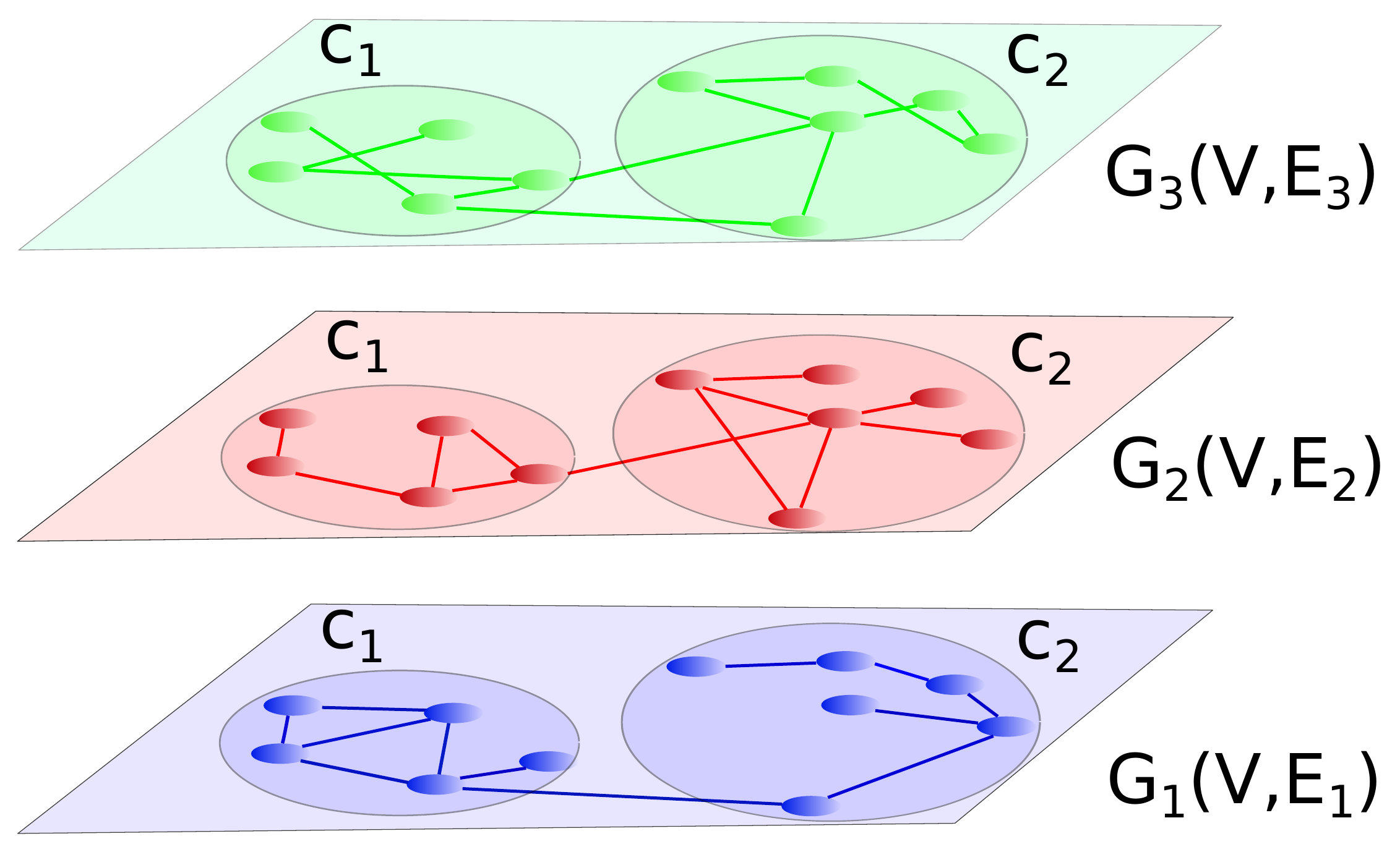}
\caption{An example of a multiplex network with 11 nodes present in  three complementary layers denoted in different colors. Two different communities across all three layers can be identified.}
\label{fig:mltplx}
\end{figure}

However, many real-world systems are naturally represented with multiple types of relationships, or with relationships that chance in time. Such systems include subsystems or layers of connectivity representing different modes of complexity. For instance, in social systems, users in social networks engage in different types of interactions (e.g., personal, professional, social, etc.). In biology,  different experiments or measurements can provide different types of interactions between genes. Reducing these networks to a single type interactions by disregarding their multiple modalities is often a very crude approximation that fails to capture a rich and holistic complexity of the system. In order to encompass a multimodal nature of these relations, a {\em multiplex network} representation has been proposed \cite{Kivela2014}. Multiplex networks (also known as {\em multidimensional}, {\em multiview} or {\em multilayer} networks) have recently attracted a lot of attention in network science community. They can be represented as a set of graph layers that share a common set of nodes, but different set of edges in each layer (cf. Fig. \ref{fig:mltplx}). With the emergence of this network representation, finding composite communities across different layers of multiplex network has become a new challenge \cite{Kim2015,Kivela2014}. 

Here, distinct from the previous approaches (reviewed in Section \ref{sec:background}), we focus on \textit{multiplex community detection}. Concretely, a novel and general model, namely the {\bf N}etwork {\bf F}usion for {\bf C}omposite {\bf C}ommunity {\bf E}xtraction (NF-CCE) along with its algorithmic framework is developed in Section \ref{sec:model}. The heart of the  NF-CCE is the Collective Non-negative Matrix Factorization (CNMF), which is employed in order to collectively factorize adjacency matrices representing  different layers in the network. The collective factorization facilitate us to obtain a consensus low-dimensional latent representation, shared across the decomposition, and hence to reveal the communities shared between the network layers. The contributions of the paper are as follows:

\begin{enumerate}

\item Inspired by recent advances in non-negative matrix factorization (NMF) techniques for graph clustering \cite{Kuang2012,Wang2011} and by using tools for subspace analysis on Grassmann manifold \cite{Dong2014,Amari1998,Panagakis2010}, we propose a general framework for extracting composite communities from multiplex networks. In particular, the framework NF-CCE, utilizes four different NMF techniques, each of which is generalized for collective factorization of adjacency matrices representing network layers, and u for computing a consensus low-dimensional latent feature matrix shared across the decomposition that is used for extracting latent communities common to all network layers. To this end, a general model involving factorization of networks' adjacency matrices and fusion of their low-dimensional subspace representation on Grassmann manifold is proposed in Section \ref{sec:model}.

\item Unlike a few matrix factorization-based methods for multiplex community extraction that have been proposed so far, e.g., \cite{Cheng2013,Tang2009,Dong2012,Panagakis2010}, that directly decompose matrices representing network layers into a low-dimensional representation common to all network layers, NF-CCE is conceptually different. Namely, it works in two steps: first, it denoises each network layer by computing its non-negative low-dimensional representation. Then it merges the low-dimensional representations into a consensus low-dimensional representation common to all network layers. This makes our method more robust to noise, and consequently, it yields much stable clustering results.   

\item Four efficient algorithms based on four different NMF techniques are developed for NF-CCE using the concept of {\em natural gradient} \cite{Amari1998,Panagakis2010} and presented in the form of {\em multiplicative update rules} \cite{LeeSeung2000} in Section \ref{sec:model}.

\end{enumerate}

The advantages of the NF-CCE over the state-of-the-art in community detection are demonstrated by conducting extensive experiments on a wide range of real-world multiplex networks, including biological, social, economic, citation, phone communication, and brain multiplex networks. In particular, we compared the clustering performance of our four methods with 6 state-of-the-art methods and 5 baseline methods (i.e., single-layer methods modified for multiplex networks), on 9 different real-world multiplex networks including 3 large-scale multiplex biological networks of 3 different species. Experimental results, in Section~\ref{sec:results}, indicate that the proposed methods yield much stable clustering results than the {\em state-of-the-art} and baseline methods, by robustly handling incomplete and noisy network layers. The experiments conducted on multiplex biological networks of 3 different species indicate NF-CCE as a superior method for finding composite communities (in biological networks also known as {\em functional modules}). Moreover, the results also indicate that NF-CCE can extract unique and more functionally consistent modules by considering all network layers together than by considering each network layer separately.

\paragraph*{{\em Notations}} throughout the paper, matrices are
denoted by uppercase boldface letters, e.g., $\mathbf{X}$. Subscript indices denote matrix elements, e.g., $\mathbf{X}_{ij}$, whereas superscript indices in brackets denote network layer, e.g., $\mathbf{X}^{(i)}$. The set of real numbers is denoted by  $\mathbb{R}$. $\vert \cdot \vert$ denotes the cardinality of a set, e.g., $\vert S \vert$. A binary matrix of size $n \times m$ is represented by $\{0, 1\}^{n \times m}$.

\section{Background and related work}\label{sec:background}

\subsection{Single-layer (monoplex) networks}

In graph theory, a monoplex network (graph) can be represented as an ordered pair, $G = (V, E)$, where $V$ is a set of $n = |V|$ vertices or nodes, and $E$ is a set of $m = |E|$ edges or links between the vertices \cite{West2001book}.  An adjacency matrix representing a graph $G$ is denoted by  $\mathbf{A} \in \{0, 1\}^{n \times n}$, where $\mathbf{A}_{ij} = 1$ if there is an edge between vertices $i$ and $j$, and $\mathbf{A}_{ij} = 0$ otherwise. Most of the real-world networks that we consider throughout the paper are represented as {\em edge-weighted graphs}, $G = (V, E, w)$, where $w: E \to \mathbb{R}$ assigns real values to edges. In this case, the adjacency matrix  instead of being a binary matrix, is a real one i.e., $\mathbf{A} \in \mathbb{R}^{n \times n}$, with entries characterizing the strength of association or interaction between the network nodes. 

Although, there is no universally accepted mathematical definition of the community notion in graphs, the probably most commonly accepted definition is the following: a community is a set of nodes in a network that are connected more densely among each other than they are to the rest of the network \cite{Kivela2014}. Hence, the problem of community detection is as follows: given an adjacency matrix $\mathbf{A}$ of one network with $n$ nodes and $k$ communities, find the community assignment of all nodes, denoted by $\mathbf{H} \in \{0, 1\}^{n \times k}$, where $\mathbf{H}_{ij} = 1$ if nodes $i$ belongs to community $j$, and $\mathbf{H}_{ij} = 0$ otherwise. We consider the case of {\em non-overlapping} communities, where a node can belong to only one community, i.e., $\sum_{j=1}^k\mathbf{H}_{ij} = 1$.

To address the community detection problem in monoplex networks, several methods have been proposed. Comprehensive surveys of these methods are  \cite{Fortunato2010} and \cite{Schaeffer2007}. To make the paper self-contained, here, we briefly review some of the most representative approaches, including {\em graph partitioning, spectral clustering, hierarchical clustering, modularity maximization, statistical inference and structure-based methods}, as well as method that rely on {\em non-negative matrix factorizations}:

\begin{itemize}

\item \textit{Graph partitioning} aims to group nodes into partitions such that the {\em cut size}, i.e., the total number of edges between 
any two partitions, is minimal. Two widely used graph partitioning algorithms that also take into account the size of partitions 
are {\em Ratio Cut} and {\em Normalized Cut} \cite{Shi2000}. Graph partitioning algorithms can be alternatively defined as spectral algorithms in which the objective is to partition the nodes into communities based on their eigenvectors obtained from {\em eigendecomposition of graph Laplacian matrix} \cite{vonLuxburg2007}. 

\item In \textit{ hierarchical clustering} the goal is to reveal network communities and their hierarchical structure based on a similarity (usually topological) measure computed between pairs of nodes \cite{Hastie_book}.
 
\item \textit{Modularity-based} algorithms are among the most popular ones. Modularity was designed to measure the strength of partition of a network into communities. It is defined as a fraction of the edges that fall within the community minus the expected fraction when these edges are randomly distributed  \cite{Girvan2002,Newman2006}. Various algorithms have been proposed for modularity optimization, including greedy techniques, simulated annealing, spectral optimization, etc. \cite{Fortunato2010}.

\item  \textit{Statistical inference} methods aims at fitting the {\em generative model} to the network data based on some hypothesis. Most commonly used statistical inference method for community detection is
the {\em stochastic  block model}, that aims to approximate a given adjacency matrix by a block structure \cite{Holland1983}. Each block in the model represents a community.

\item  \textit{Structure-based} methods aim to find subgraphs representing meta definitions of communities. Their objective is to find maximal cliques, i.e., the cliques which are not the subgraph of any other clique. The union of these cliques form a subgraph, whose components are interpreted as communities \cite{Everett1998}. 

\item More recently, graph clustering methods that rely on the \textit{Non-Negative Matrix Factorization} (NMF) \cite{LeeSeung2000}
have been proposed e.g., \cite{Kuang2012, Wang2011}. Their goal is to approximate a symmetric  adjacency matrix  of a given network by a product of two non-negative, low-rank matrices, such that they have clustering interpretation, i.e., they can be used for assigning nodes to communities. The proposed methods here, follow this line of research, but as opposed to the existing methods \cite{Kuang2012, Wang2011}, the NF-CCE can effectively handle multiplex networks.

\end{itemize}

\subsection{Multiplex networks}
\label{sec:mltplx}

A multiplex network is a set of $N$ monoplex networks (or layers), $G_i (V, E_i)$, for $i = 1, \dots, N$. The number of nodes in each layer is the same, $n = |V|$, while the connectivity pattern and the distribution of links in each layer differs, $m_i = |E_i|$ (see Fig. \ref{fig:mltplx}). Similarly to monoplex networks, we consider the case where each layer represents a weighted, undirected graph, i.e., $G_i (V, E_i, w_i)$. A multiplex network can be represented as a set of adjacency matrices encoding connectivity patterns of individual layers, $\mathbf{A}^{(i)} \in \mathbb{R}^{n \times n}$, for $i = 1, \dots, N$. The goal of community detection in multiplex networks is to infer shared, latent community assignment that best fits all given layers. Given that each layer contains incomplete and complementary information, this process of finding shared communities by integrating information from all layers is also known in the literature as {\em network integration (fusion)} \cite{Zhang2015,Tang2009}.      

Unlike the case of monoplex networks, research on community detection in multiplex networks is scarce. Existing methods extract communities from multiplex networks first by aggregating the links of all layers into a single layer, and then applying a monoplex method to that single layer \cite{Berlingerio2011, Tang2009, Rodriguez2010}.  However, this approach does not account for shared information between layers and treats the noise present in each layer uniformly. Clearly, this is not the case in real-world multiplex networks where each level is contaminated by different noise in terms of magnitude and, possibly, distribution. Thus, by aggregating links from different layers the noise in the aggregated layer significantly increases, resulting in a poor community detection performance.

Current state-of-the-art methods are built on monoplex approaches and further generalized to multiplex networks. They can be divided into the following categories:

\begin{itemize}

\item {\em Modularity-based} approaches that generalize the notion of modularity from single-layer to multi-layer networks \cite{Mucha2010}. Namely, to alleviate the above mentioned limitations, the Principal Modularity Maximization (PMM) \cite{Tang2009} has been proposed. First, for each layer, PMM extracts structural features by optimizing its modularity, and thus significantly denoising each layer; then, it applies PCA on concatenated matrix of structural feature matrices, to find the principal vectors, followed by K-means to perform clustering assignment. The main drawback of this approach is that it treats structural feature matrices of all layers on equal basis (i.e., it is not capable of distinguishing between more and less informative network layers, or complementary layers). Even though the noise is properly handled  by this method, the {\em complementarity aspect} cannot be captured well by the integration step.

\item {\em Spectral clustering} approaches that generalize the eigendecomposition from single to multiple Laplacian matrices representing network layers. One of the {\em state-of-the-art} spectral clustering methods for multiplex graphs is the Spectral Clustering on Multi-Layer (SC-ML) \cite{Dong2014}. First, for each network layer, SC-ML computes a subspace spanned by the principal eigenvectors of its Laplacian matrix. Then, by interpreting each subspace as a point on Grassmann manifold, SC-ML merges subspaces into a consensus subspace from which the composite clusters are extracted. The biggest drawback of this methods is the underlying spectral clustering, that always finds tight and small-scale and, in some cases, almost trivial communities. For example, SC-ML cannot adequately handle network layers with missing or weak connections, or layers that have disconnected parts.

\item {\em Information diffusion-based} approaches that utilize the concept of diffusion on networks to integrate network layers. One of such methods is Similarity Network Fusion (SNF) proposed by Wang {\em et al.} \cite{Wang2014}. SNF captures both shared and complementary information in network layers. It computes a fused matrix from the similarity matrices derived from all layers through parallel interchanging diffusion process on network layers. Then, by applying a spectral clustering method on the fused matrix they extract communities. However, for sparse networks, the diffusion process, i.e., information propagation, is not very efficient and it may results in poor clustering performance.

\item {\em Matrix and tensor factorization-based} approaches that utilize collective factorization of adjacency matrices representing network layers. A few matrix and tensor decomposition-based approaches have been proposed so far \cite{Tang2009lmf,Dong2012,Papalexakis2013,Cheng2013}. Tang {\em et al.} \cite{Tang2009lmf} introduced the  Linked Matrix Factorization (LMF) which fuses information from multiple network layers by factorizing each adjacency matrix into a layer-specific factor and a factor that is common to all network layers. Dong {\em et al.} \cite{Dong2012}, introduced the Spectral Clustering with Generalized Eigendecomposition (SC-GED) which factorizes Laplacian matrices instead of adjacency matrices. Papalexakis {\em et al.} \cite{Papalexakis2013} proposed GraphFuse, a method for clustering multi-layer networks based on sparse PARAllel FACtor (PARAFAC) decomposition \cite{Harshman1970} with non-negativity constraints. Cheng {\em et al.} introduced Co-regularized Graph Clustering based on NMF (CGC-NMF). They factorize each adjacency matrix using symmetric NMF while keeping the Euclidean distance between their non-negative low-dimensional representations small. As already pointed out in Section \ref{sec:introduction}, one of the major limitations of all of these factorization methods is that they treat each network layer on an equal basis and, unlike PMM or SC-ML, for example, they cannot filter out irrelevant information or noise. 

\end{itemize}

To alleviate the drawbacks of the aforementioned methods, the NF-CCE framework is detailed in the following section. It consists of $4$ models, where each layer is first denoised by computing its non-negative low-dimensional subspace representation. Then, the low-dimensional subspaces are merged into a consensus  subspace whose non-negative property enables clustering interpretation. The models are conceptually similar to the SC-ML method, since they use the same merging technique to find the common subspace of all layers.

\section{Proposed Framework}\label{sec:model}

Here, we present four novel metods that are built upon $4$ non-negative matrix factorization models, SNMF \cite{Kuang2012}, PNMF \cite{Yang2010}, SNMTF \cite{Ding2006} and Semi-NMTF \cite{Ding2010}, and extended for fusion and clustering of multiplex networks. Since the derivation of each method is similar, we present them in a unified framework, namely NF-CCE. NF-CCE extracts composite communities from a multiplex network consisting of $N$ layers. In particular, given $N$-layered multiplex network represented by adjacency matrices, $\{\mathbf{A}^{(1)}, \dots, \mathbf{A}^{(N)}  \}$, NF-CCE consists of two steps: 

\paragraph*{\textit{Step 1}} For each network layer, $i$, we obtain its non-negative, low-dimensional representation, $\mathbf{H}^{(i)}$, under column orthonormality constraints i.e., $\mathbf{H}^{(i)T}\mathbf{H}^{(i)} = \mathbf{I}$, by using any of the  non-negative factorization methods mentioned above.

\paragraph*{\textit{Step 2}} We fuse the low-dimensional representations into a common, consensus representation, $\mathbf{H}$, by proposing a collective matrix factorization model. That is, we collectively decompose all adjacency matrices, 
$\mathbf{A}^{(i)}$ into a common matrix, $\mathbf{H}$,  whilst enforcing the non-negative low-dimensional representation of network layers, $\mathbf{H}^{(i)}$ (computed in the previous step), to be close enough to the consensus low-dimensional representation, $\mathbf{H}$. The general objective function capturing these two properties is written as follows:

\begin{equation}
\min\limits_{\mathbf{H} \geq 0} \mathcal{J} = \sum_{i=1}^{N} \mathcal{J}^{(i)} (\mathbf{H}; \mathbf{A}^{(i)}) + \alpha \sum_{i=1}^{N} \mathcal{J}_{c}^{(i)} (\mathbf{H}; \mathbf{H}^{(i)}) 
\label{eq:gen}
\end{equation}

\noindent where, $\mathcal{J}^{(i)}$ is an objective function for clustering $ith$ layer and $\mathcal{J}_{c}^{(i)}$ is the loss function quantifying the inconsistency between each low-dimensional representation $\mathbf{H}^{(i)}$, computed in {\em Step 1}, and the consensus representation $\mathbf{H}$. 
%\end{itemize}

Below we provide the details of the second step for each individual factorization technique. 

\subsection{Collective SNMF (CSNMF)}\label{sec:csnmf}

We factorize each individual adjacency matrix using Symmetric NMF in the following way:

\begin{equation*}
\mathbf{A}^{(i)} \approx \mathbf{H}\mathbf{H}^{T}
\end{equation*}

\noindent under the following constraints: $\mathbf{H} \geq 0$ and $\mathbf{H}^{T}\mathbf{H} = \mathbf{I}$; where, $i = 1,\dots,N$. 

The first part of our general objective function in {\em Step 2} (Eq. \ref{eq:gen}) has the following form: 
\begin{equation}\label{eq:csnmf}
\mathcal{J}^{(i)}(\mathbf{H}; \mathbf{A}^{(i)}) =  \parallel \mathbf{A}^{(i)} - \mathbf{H}\mathbf{H}^{T} \parallel_{F}^2
\end{equation}

\noindent where $\mathbf{H}$ is the consensus, non-negative low-dimensional matrix, and $F$ denotes Frobenius norm.

\subsection{Collective PNMF (CPNMF)}\label{sec:cpnmf}
We factorize each individual adjacency matrix using Projective NMF in the following way:

\begin{equation*}
\mathbf{A}^{(i)} \approx \mathbf{H}\mathbf{H}^{T}\mathbf{A}^{(i)}
\end{equation*}

\noindent under the following constraints:  $\mathbf{H} \geq 0$ and  $\mathbf{H}^{T}\mathbf{H} = \mathbf{I}$; where, $i = 1,\dots,N$.

The first part of our general objective function in {\em Step 2} (Eq. \ref{eq:gen}) has the following form: 
\begin{equation}\label{eq:cpnmf}
\mathcal{J}^{(i)}(\mathbf{H}; \mathbf{A}^{(i)}) =  \parallel \mathbf{A}^{(i)} - \mathbf{H}\mathbf{H}^{T}\mathbf{A}^{(i)} \parallel_{F}^2
\end{equation}
\noindent where $\mathbf{H}$ is the consensus, non-negative low-dimensional matrix.

\subsection{Collective SNMTF (CSNMTF)}\label{sec:csnmtf}

We tri-factorize each individual adjacency matrix using Symmetric NMTF in the following way:

\begin{equation*}
\mathbf{A}^{(i)} \approx \mathbf{H}\mathbf{S}^{(i)} \mathbf{H}^T
\end{equation*}

\noindent under the following constraints: $\mathbf{H} \geq 0$ and  $\mathbf{H}^{T}\mathbf{H} = \mathbf{I}$; where $i = 1,\dots,N$.

The first part of our general objective function in {\em Step 2} (Eq. \ref{eq:gen}) has the following form:
\begin{equation}\label{eq:csnmtf}
\mathcal{J}^{(i)}(\mathbf{H}; \mathbf{A}^{(i)}, \mathbf{S}^{(i)}) =  \parallel \mathbf{A}^{(i)} - \mathbf{H}\mathbf{S}^{(i)}\mathbf{H}^{T} \parallel_{F}^2
\end{equation}

\noindent 
where $\mathbf{H}$ is the consensus low-dimensional matrix. 

In the derivation of our algorithm, we distinguish between two cases. In the first case, we consider $\mathbf{S}$ matrix to be non-negative, i.e., $\mathbf{S}^{(i)} \geq 0$. We call that case CSNMTF. In the second case, we consider  $\mathbf{S}^{(i)}$ matrix to have both positive and negative entries. We call this case collective symmetric semi-NMTF, or CSsemi-NMTF (or CSsNMTF).

\subsection{Merging low-dimensional representation of graph layers on Grassmann Manifolds}
\label{sec:grassmann}

For the second term in our general objective function (Eq. \ref{eq:gen}) in {\em Step 2}, we utilize the orthonormal property of non-negative, low-dimensional matrices, $\mathbf{H}^{(i)}$, and propose a distance measure based on this property. Namely, Dong {\em et al.} \cite{Dong2014} proposed to use the tools from subspace analysis on Grassmann manifold. A Grassmann manifold $\mathbb{G}(k,n)$ is a set of $k$-dimensional linear subspaces in $\mathbb{R}^{n}$ \cite{Dong2014}. Given that, each orthonormal cluster indicator matrix, $\mathbf{H}_{i} \in \mathbb{R}^{n \times k}$, spanning the corresponding $k$-dimensional non-negative subspace, $span(\mathbf{H}_{i})$ in $\mathbb{R}^{n}$, is mapped to a unique point on the Grassmann manifold $\mathbb{G}(k,n)$. The geodesic distance between two subspaces, can be computed by projection distance. For example, the square distance between two subspaces, $\mathbf{H}_i$ and $\mathbf{H}_j$, can be computed as follows:

\begin{align*}
d^2_{proj} (\mathbf{H}_{i}, \mathbf{H}_{j}) & = \sum_{i=1}^{k} sin^2\theta_i = k - \sum_{i=1}^{k} cos^2\theta_i\\
& = k -tr(\mathbf{H}_{i}\mathbf{H}_{i}^T
\mathbf{H}_{j}\mathbf{H}_{j}^T)\\
\end{align*}

To find a consensus subspace, $\mathbf{H}$, we factorize all the adjacency matrices, $\mathbf{A}^{(i)}$, and minimize the distance between their subspaces and the consensus subspace on Grassmann manifold. Following this approach, we can write the second part of our general objective function in the following way:

\begin{align}\label{eq:distance}
\mathcal{J}_{c}^{(i)}(\mathbf{H}, \mathbf{H}^{(i)}) & = k - tr(\mathbf{H}\mathbf{H}^{T}\mathbf{H}^{(i)}\mathbf{H}^{(i)T})\\ \nonumber
& = \parallel \mathbf{H}\mathbf{H}^T - \mathbf{H}^{(i)}\mathbf{H}^{(i)T} \parallel_F^2
\end{align}

\subsection{Derivation of the general multiplicative update rule}

In {\em Step 1}, we use well-known non-negative factorization techniques, namely SNMF, PNMF, SNMTF and Ssemi-NMTF, for which the update rules for computing low-dimensional non-negative matrices, $\mathbf{H}^{(i)}$, have been provided in the corresponding papers \cite{Kuang2012,Yang2010,Ding2006,Ding2010}, respectively. They are summarized in Table \ref{tab:mur}. As for the {\em Step 2}, we derive the update rules for each of the collective factorization techniques presented in Sections \ref{sec:csnmf}, \ref{sec:cpnmf} and \ref{sec:csnmtf}. The details of the derivation are given in Section 2 of the online supplementary material. Here we provide a general update rule for Equation \ref{eq:gen}.

We minimize the general objective function shown in Equation \ref{eq:gen}, under the following constraints: $\mathbf{H} \geq 0$  and $\mathbf{H}^T\mathbf{H} = \mathbf{I}$. Namely, we adopt the idea from Ding {\em et al.}\cite{Ding2006} to impose orthonormality constraint on $\mathbf{H}$ matrix, i.e., $\mathbf{H}^T\mathbf{H} = \mathbf{I}$; that has been shown to lead to a more rigorous clustering interpretation \cite{Ding2006}. Moreover, assignments of network nodes to composite communities can readily be done by examining the entries in rows of $\mathbf{H}$ matrix. Namely, we can interpret matrix $\mathbf{H}^{n \times k}$ as the {\em cluster indicator matrix}, where the entries in $i$-th row (after row normalization) can be interpreted as a posterior probability that a node $i$ belongs to each of the $k$ composite communities. In all our experiments, we apply {\em hard clustering} procedure, where  a node is assign to the cluster that has the largest probability value.

We derive the update rule, for matrix $\mathbf{H}$, for minimizing the objective function (Eq. \ref{eq:gen}) following the procedure from the constrained optimization theory \cite{Boyd2004}. Specifically, we follow the strategy employed in the derivation of NMF \cite{LeeSeung2000} to obtain a multiplicative update rule for $\mathbf{H}$ matrix that can be used for finding a local minimum of the optimization problem (Eq. \ref{eq:gen}).

The derivative of the objective function (Eq. \ref{eq:gen}) with respect to $\mathbf{H}$ is as follows:

\begin{equation}\label{eq:ordinary}
\nabla_{\mathbf{H}}\mathcal{J} =  \sum\limits_{i=1}^N \nabla_{\mathbf{H}}\mathcal{J}^{(i)}(\mathbf{H}; \mathbf{A}^{(i)}) - \alpha\sum\limits_{i=1}^N \mathbf{H}^{(i)}\mathbf{H}^{(i)T}\mathbf{H}
\end{equation}

\noindent where the first term under summation can be decomposed into two non-negative terms, namely:
\begin{equation*}
\small
%\sum\limits_{i=1}^N 
\nabla_{\mathbf{H}}\mathcal{J}^{(i)}(\mathbf{H}; \mathbf{A}^{(i)}) = 
%\sum\limits_{i=1}^N 
[\nabla_{\mathbf{H}}\mathcal{J}^{(i)}(\mathbf{H}; \mathbf{A}^{(i)})]^{+} - %\sum\limits_{i=1}^N 
[\nabla_{\mathbf{H}}\mathcal{J}^{(i)}(\mathbf{H}; \mathbf{A}^{(i)})]^{-}
\end{equation*}

\noindent where, $[\nabla_{\mathbf{H}}\mathcal{J}^{(i)}(\mathbf{H}; \mathbf{A}^{(i)})]^{+} \geq 0$, $[\nabla_{\mathbf{H}}\mathcal{J}^{(i)}(\mathbf{H}; \mathbf{A}^{(i)})]^{-} \geq 0$ are non-negative terms. Depending on the type of collective factorization technique represented in Section \ref{sec:csnmf}, \ref{sec:cpnmf} or \ref{sec:csnmtf}, the first term represents the derivative of the corresponding objective function, i.e., Equation \ref{eq:csnmf}, \ref{eq:cpnmf} or \ref{eq:csnmtf}, respectively. The second term represents a derivative of Equation \ref{eq:distance} with respect to $\mathbf{H}$ . 

To incorporate the orthonormality constraint into the update rule, we introduce the concept of {\em natural gradient} by following the work of Panagakis {\em et al.} \cite{Panagakis2010}. Namely, we shown in Section \ref{sec:grassmann} that columns of $\mathbf{H}$ matrix span a vector subspace known as Grassmann manifold $\mathbb{G}(k,n)$, i.e., $span(\mathbf{H}) \in \mathbb{G}(k,n)$ \cite{Panagakis2010}. Using that, Amari in \cite{Amari1998} has showed that when an optimization problem is defined over a Grassmann manifold, the ordinary gradient of the optimization function (Equation \ref{eq:ordinary}) does not represent its steepest direction, but {\em natural gradient} does \cite{Amari1998}. 

Therefore, we define a natural gradient to optimize our objective function (\ref{eq:gen}) under the orthornormality constraint. Following Panagakis {\em et al.} \cite{Panagakis2010}, the natural gradient of $\mathcal{J}$ on Grassmann manifold at $\mathbf{H}$ can be written in terms of the ordinary gradient as follows:

\begin{equation}\label{eq:gradient}
\widetilde{\nabla}_{\mathbf{H}}\mathcal{J} = 
\nabla_{\mathbf{H}} \mathcal{J} - \mathbf{H}\mathbf{H}^T\nabla_{\mathbf{H}} \mathcal{J} 
\end{equation}

\noindent where, $\nabla_{\mathbf{H}} \mathcal{J}$ is the ordinary gradient given in Equation \ref{eq:ordinary}.

Following the Karush-Kuhn-Tucker (KKT) complementarity condition \cite{Boyd2004} and preserving the non-negativity of $\mathbf{H}$, the general update rule for $\mathbf{H}$ matrix using the natural gradient is as follows:
\begin{equation}\label{eq:mul}
\mathbf{H}_{jk} \leftarrow \mathbf{H}_{jk} \circ
\frac{[\widetilde{\nabla}_{\mathbf{H}}\mathcal{J}]^{-}_{jk}}{[\widetilde{\nabla}_{\mathbf{H}}\mathcal{J}]^{+}_{jk}}
\end{equation}

\noindent where, ``$\circ$'' denotes Hadamard product. 

\begin{align*}
[\widetilde{\nabla}_{\mathbf{H}}\mathcal{J}]^{-} & = \mathbf{H}\mathbf{H}^T \sum\limits_{i=1}^N [\nabla_{\mathbf{H}}\mathcal{J}^{(i)}(\mathbf{H}; \mathbf{A}^{(i)})]^{+} \\
& + \sum\limits_{i=1}^N [\nabla_{\mathbf{H}}\mathcal{J}^{(i)}(\mathbf{H}; \mathbf{A}^{(i)})]^{-} + \alpha\sum\limits_{i=1}^N \mathbf{H}^{(i)}\mathbf{H}^{(i)T}\mathbf{H}
\end{align*}

\begin{align*}
[\widetilde{\nabla}_{\mathbf{H}}\mathcal{J}]^{+} & = \mathbf{H}\mathbf{H}^T \sum\limits_{i=1}^N [\nabla_{\mathbf{H}}\mathcal{J}^{(i)}(\mathbf{H}; \mathbf{A}^{(i)})]^{-}\\ 
& + \sum\limits_{i=1}^N [\nabla_{\mathbf{H}}\mathcal{J}^{(i)}(\mathbf{H}; \mathbf{A}^{(i)})]^{+} + \alpha \mathbf{H}\mathbf{H}^T
\sum\limits_{i=1}^N \mathbf{H}^{(i)}\mathbf{H}^{(i)T}\mathbf{H}
\end{align*}

The concrete update rule for each collective factorization method is summarized in Table \ref{tab:mur} and united within {\em NF-CCE} algorithm (see Algorithm \ref{alg:alg1}). 
%Regarding the correctness and convergence of the Algorithm 1 please refer to the appendix.

\begin{algorithm}\label{alg:NF-CCE}
\caption{NF-CCE\newline 
    \textbf{Input}: Adjacency  matrices $\mathbf{A}^{(i)}$ for each network layer $i = 1, \dots, N$; number of clusters $k$; parameter $\alpha$; factorization technique: {\small FACTORIZATION} $\in$ \{{\small SNMF}, {\small PNMF}, {\small SNMTF}\} \newline
    \textbf{Output}: Consensus cluster indicator matrix $\mathbf{H}$}
\label{alg:alg1}
\begin{algorithmic}\footnotesize
\SWITCH {FACTORIZATION}
	\CASE {'SNMF'}
		\FOR{$ i \in [1, N]$}
		\STATE $\mathbf{H}^{(i)} \leftarrow \text{FACTORIZATION}(\mathbf{A}^{(i)}, k)$
        \ENDFOR
        \STATE $\mathbf{A}_{avg} \leftarrow \sum\limits_{i=1}^N \mathbf{A}^{(i)} + 	   \frac{\alpha}{2} \mathbf{H}^{(i)}\mathbf{H}^{(i)T}$
        \STATE $\mathbf{H} \leftarrow \text{FACTORIZATION}(\mathbf{A}_{avg}, k)$
	\ENDCASE
    \CASE {'PNMF'}
		\FOR{$ i \in [1, N]$}
		\STATE $\mathbf{H}^{(i)} \leftarrow \text{FACTORIZATION}(\mathbf{A}^{(i)}, k)$
        \ENDFOR
        \STATE $\mathbf{A}_{avg} \leftarrow \sum\limits_{i=1}^N \mathbf{A}^{(i)}\mathbf{A}^{(i)T}  + \alpha \mathbf{H}^{(i)}\mathbf{H}^{(i)T}$
        \STATE $\mathbf{H} \leftarrow \text{FACTORIZATION}(\mathbf{A}_{avg}, k)$
	\ENDCASE
    \CASE {'SNMTF'}
		\FOR{$ i \in [1, N]$}
		\STATE ($\mathbf{H}^{(i)}, \mathbf{S}^{(i)}) \leftarrow 
        \text{FACTORIZATION}(\mathbf{A}^{(i)}, k)$
        \ENDFOR
        \STATE $\mathbf{H} \leftarrow \text{CSNMTF}(\{\mathbf{A}^{(i)} \}_{i=1}^{N}, \{\mathbf{H}^{(i)}\}_{i=1}^N, \{\mathbf{S}^{(i)}\}_{i=1}^N, k)$
	\ENDCASE
\ENDSWITCH
\end{algorithmic}
\end{algorithm}

\begin{table*}
\begin{center}
\caption{Multiplicative update rules (MUR) for single-layer and multiplex network analysis.}
\label{tab:mur}
\resizebox{\linewidth}{!}{%
\begin{tabular}{l|c|c}
\hline
Method & Single-layer MUR & Multiplex MUR \\
\hline
SNMF & $\mathbf{H}^{(i)}_{jk} \leftarrow \mathbf{H}^{(i)}_{jk} \circ  \frac{\big [\mathbf{A}^{(i)}\mathbf{H}^{(i)}\big ]_{jk}}{\big [
\mathbf{H}^{(i)}\mathbf{H}^{(i)T}\mathbf{A}^{(i)}\mathbf{H}^{(i)} \big ]_{jk}}$ & $\mathbf{H}_{jk} \leftarrow \mathbf{H}_{jk} \circ \frac{\big [\big (\sum\limits_{i=1}^{N}\mathbf{A}^{(i)} + \frac{\alpha}{2}\mathbf{H}^{(i)}\mathbf{H}^{(i)T}\big )\mathbf{H}\big ]_{jk}}{\big [\mathbf{H}\mathbf{H}^{T}\big (\sum\limits_{i=1}^{N} \mathbf{A}^{(i)} + \frac{\alpha}{2}\mathbf{H}^{(i)}\mathbf{H}^{(i)T}\big )\mathbf{H} \big ]_{jk}}$ \\
\hline
PNMF & $\mathbf{H}^{(i)}_{jk} \leftarrow \mathbf{H}^{(i)}_{jk} \circ  \frac{\big [\mathbf{A}^{(i)}\mathbf{A}^{(i)T}\mathbf{H}^{(i)}\big ]_{jk}}{\big [
\mathbf{H}^{(i)}\mathbf{H}^{(i)T}\mathbf{A}^{(i)}\mathbf{A}^{(i)T}\mathbf{H}^{(i)} \big ]_{jk}}$ & $\mathbf{H}_{jk} \leftarrow \mathbf{H}_{jk} \circ \frac{\big [\big ( \sum\limits_{i=1}^{N} \mathbf{A}^{(i)}\mathbf{A}^{(i)T} + \alpha  \mathbf{H}^{(i)}\mathbf{H}^{(i)T}\big )\mathbf{H}\big ]_{jk}}{\big [\mathbf{H}\mathbf{H}^{T}\big (\sum\limits_{i=1}^{N} \mathbf{A}^{(i)}\mathbf{A}^{(i)T} + \alpha\mathbf{H}^{(i)}\mathbf{H}^{(i)T}\big )\mathbf{H} \big ]_{jk}}$ \\
\hline
\multirow{2}{*}{\vspace{-5mm} SNMTF} & $\mathbf{H}^{(i)}_{jk} \leftarrow \mathbf{H}^{(i)}_{jk} \circ \frac{\big [ \mathbf{A}^{(i)}\mathbf{H}^{(i)}\mathbf{S}^{(i)}\big ]_{jk}}{\big [\mathbf{H}^{(i)}\mathbf{H}^{(i)T}\mathbf{A}^{(i)}\mathbf{H}^{(i)}\mathbf{S}^{(i)}\big ]_{jk}}$ & \multirow{2}{*}{ $\mathbf{H}_{jk} \leftarrow \mathbf{H}_{jk} \circ \frac{\big [  \sum\limits_{i=1}^{N} \mathbf{A}^{(i)}\mathbf{H}\mathbf{S}^{(i)} + \frac{\alpha}{2} \sum\limits_{i=1}^N \mathbf{H}^{(i)}\mathbf{H}^{(i)T}\mathbf{H}  \big ]_{jk}}{\big [ \mathbf{H}\mathbf{H}^T \big (\sum\limits_{i=1}^{N} \mathbf{A}^{(i)}\mathbf{H}\mathbf{S}^{(i)} + \frac{\alpha}{2} \sum\limits_{i=1}^N \mathbf{H}^{(i)}\mathbf{H}^{(i)T}\mathbf{H} \big ) \big ]_{jk}}$ } \\
 & $\mathbf{S}^{(i)}_{jk} \leftarrow \mathbf{S}^{(i)}_{jk} \circ \frac{\big [ \mathbf{H}^{(i)T}\mathbf{A}^{(i)}\mathbf{H}^{(i)} \big]_{jk}}{\big [\mathbf{H}^{(i)T}\mathbf{H}^{(i)}\mathbf{S}^{(i)}\mathbf{H}^{(i)T}\mathbf{H}^{(i)} \big ]_{jk}}$ & \\
\hline
\multirow{2}{*}{\vspace{-4mm} SsNMTF} & $\mathbf{H}^{(i)}_{jk} \leftarrow \mathbf{H}^{(i)}_{jk} \circ \frac{ \Big [ \big [ \mathbf{A}^{(i)}\mathbf{H}^{(i)}\mathbf{S}^{(i)} \big ]^{+} + \mathbf{H}^{(i)}\mathbf{H}^{(i)T} [ \mathbf{A}^{(i)}\mathbf{H}^{(i)}\mathbf{S}^{(i)} \big ]^{-}  \Big ]_{jk} }{\Big [ \big [ \mathbf{A}^{(i)}\mathbf{H}^{(i)}\mathbf{S}^{(i)} \big ]^{-} + \mathbf{H}^{(i)}\mathbf{H}^{(i)T} [ \mathbf{A}^{(i)}\mathbf{H}^{(i)}\mathbf{S}^{(i)} \big ]^{+}   \Big ]_{jk}}$ & 
\multirow{2}{*}{
$\mathbf{H}_{jk} \leftarrow \mathbf{H}_{jk} \frac{\Big [ \sum\limits_{i=1}^{N} \big [ \mathbf{A}^{(i)}\mathbf{H}\mathbf{S}^{(i)} \big ]^{+} + \mathbf{H}\mathbf{H}^T \big [ \mathbf{A}^{(i)}\mathbf{H}\mathbf{S}^{(i)} \big ]^{-} + \frac{\alpha}{2} \mathbf{H}^{(i)}\mathbf{H}^{(i)T}\mathbf{H} \Big ]_{jk}}{\Big [ \sum\limits_{i=1}^{N} \big [ \mathbf{A}^{(i)}\mathbf{H}\mathbf{S}^{(i)} \big ]^{-} + \mathbf{H}\mathbf{H}^{T}\Big ( \big [ \mathbf{A}^{(i)}\mathbf{H}\mathbf{S}^{(i)} \big ]^{+} + \frac{\alpha}{2}\mathbf{H}^{(i)}\mathbf{H}^{(i)T}\mathbf{H} \Big )\Big ]_{jk}}$ }\\
& $\mathbf{S}^{(i)} \leftarrow \big ( \mathbf{H}^{(i)T}\mathbf{H}^{(i)} \big )^{-1}
\mathbf{H}^{(i)T}\mathbf{A}^{(i)}\mathbf{H}^{(i)}(\mathbf{H}^{(i)T}\mathbf{H}^{(i)} \big )^{-1} $ & \\
\hline
\end{tabular}}
\end{center}
\end{table*}

\section{Experiments}\label{sec:exp}

We test our methods on synthetic, as well as on real-world data. We designed synthetic multiplex networks with clear ground truth information and different properties in terms of noise and complementary information of network layers. The goal is to address the robustness of our methods against noise and their ability to handle complementary information contained in layers. The real-world multiplex networks are taken from diverse experimental studies to demonstrate the applicability of our methods in a broad spectrum of disciplines. Namely, we consider social and biological networks, networks of mobile phone communications, brain networks and networks constructed from bibliographic data. The biological networks are treated as a special case because of their lack of ground truth clusters. We provide detailed analysis of such networks based on the functional annotations of their nodes (genes). We present results of comparative analysis of our proposed methods against {\em state-of-the-art} methods described in Section \ref{sec:mltplx}. Specifically, we compare our methods against PMM, SC-ML, SNF, LMF, GraphFuse and CGC-NMF. Moreover, we adopt the following single-layer methods, SNMF, SNMTF, PNMF and MM (modularity maximization) to be our baseline methods. 

\subsection{Synthetic multiplex networks}

We generate two sets of synthetic multiplex networks. First type, that we denote {\em SYNTH-C}, is designed to demonstrate complementary information in layers; whereas the second type, that we denote {\em SYNTH-N}, is designed to demonstrate different levels of noise between communities contained in layers. Our synthetic networks are generated by using {\em planted partition model} \cite{Condon2001}. The procedure is as follows: we choose the total number of nodes $n$ partitioned into $N$ communities of equal or different sizes. For each layer, we split the corresponding adjacency matrix into blocks defined by the partition. Entries in each diagonal block, are filled with ones randomly, with probability $p_{ii}$, representing the {\em within-community probability} or also referred as community edge density. We also add random noise between each pair of blocks, $ij$, with probability $p_{ij}$, representing {\em between-community probability}. The larger the values of $p_{ij}$ are the harder the clustering is. Similarly, the smaller the values of $p_{ii}$ are the harder the clustering is. We vary these probabilities across the layers to simulate complementary information and noise in the following way:

{\parindent5mm\textit{SYNTH-C}.} We generate two-layer multiplex networks with $n = 200$ nodes and $N = 2$ communities with equal number of nodes each. We generate 11 different multiplex networks with different amounts of information between two layers. Namely, we vary the within-community probability $p_{11} = \{0.05, 0.075, \dots, 0.3\}$ of the first community of the first layer across different multiplex networks, while  fixing the within-community probability of the second community, $p_{22} = 0.2$. In the second layer, we represent the complementary information by fixing the within-community probability of the first community to $p_{11} = 0.2$ and varying within-cluster probability of the second community $p_{22} = \{0.05, 0.075, \dots, 0.3\}$ across the multiplex networks. For all multiplex networks, we set the same amount of nosy links, by fixing between-community probability to $p_{12} = 0.05$. 

{\parindent5mm\textit{SYNTH-N}.} Similar to {\em SYNTH-C} we generate two-layer multiplex networks with two communities ($n$ and $N$ are the same as in {\em SYNTH-C}). We fix the within-community probability of both communities and both layers to $p_{11} = 0.3$ and $p_{22} = 0.3$ across all multiplex networks. We vary the between-community probability $p_{12} = \{0.02, 0.04, \dots, 0.2\}$ of the first layer, while keeping the between-community probability of the second layer fixed, $p_{12} = 0.02$, across all multiplex networks. 

The spy plots of adjacency matrices representing layers of {\em SYNTH-C} and {\em SYNTH-N} are given in the Section 1 of the online supplementary material.

\subsection{Real-world multiplex networks}

Below we provide a brief description of real-world multiplex networks used in our comparative study:

{\parindent5mm\textit{Bibliographic data, CiteSeer}}: the data are adopted from \cite{cora}. The network consist of $N = 3,312$ papers belonging to $6$ different research categories, that we grouped into $k = 3$ pairs categories. We consider these categories as the ground  truth classes. We construct two layers: citation layer, representing the  citation relations between papers extracted from the paper citation records; and the  paper similarity layer, constructed  by  extracting a vector of $3,703$ most frequent and unique words for each paper, and then computing the cosine similarity between each pair of papers. We construct the k-nearest neighbor graph from the similarity matrix by connecting each paper with its $10$ most similar papers.

{\parindent5mm\textit{Bibliographic data, CoRA}}: the data are adopted from \cite{cora}. The network consists  of $1,662$ machine  learning  papers  grouped  into $k = 3$ different research categories. Namely, Genetic Algorithms, Neural Networks and Probabilistic  Methods.  We  use  the  same  approach  as  for {\em CiteSeer} dataset to construct the citation and similarity layers.

{\parindent5mm\textit{Mobile phone data (MPD)}}: the data are adopted from \cite{Dong2012}. The network consists of $N = 3$ layers representing different mobile phone communications between $n = 87$ phone users on the MIT campus; namely, the layers represent physical location, bluetooth scans and phone calls. The ground truth clusters are known and manually annotated.

{\parindent5mm\textit{Social network data (SND)}}: the data are adopted from \cite{CLP}. The dataset represents the multiplex social network of a corporate law partnership, consisting of $N = 3$ layers having three types of edges, namely, co-work, friendship and advice. Each layer has $n = 71$ nodes representing employees in a law firm. Nodes have many attributes. We use the location of employees' offices  as well as their status in the firm as the ground truth for clusters and perform two different experiments, namely {\em SND(o)} and {\em SND(s)} respectively. 

{\parindent5mm\textit{Worm Brain Networks (WBN)}}: the data are retrieved from WormAtlas\footnote{http://www.wormatlas.org/}, i.e., from the original study of White {\em et al.} \cite{White1986}. The network consist of $n = 279$ nodes representing neurons, connected via $N = 5$ different types of links (i.e., layers), representing $5$ different types of synapse. We use neuron types as the ground truth clusters.

{\parindent5mm\textit{Word Trade Networks (WTN)}}: the data represents different types of trade relationships (import/export) among $n = 183$ countries in the world \cite{De2015}. The network consist of $339$ layers representing different products (goods). Since, for some products layers are very sparse, we retain the layers having more than $n-1$ links, which resulted in $N = $ layers. We use geographic regions (continents) of countries and economic trade categories for defining ground truth clusters\footnote{data about countries are downloaded from \url{http://unctadstat.unctad.org}}. Thus, we perform experiments with this two ground truth clusters, namely {\em WTN (reg)}, denoting geographic regions and {\em WTN (cat)}, denoting economic categories. 

% {\parindent5mm\textit{London Transportation Network (LTN)}}: the data are adopted from \cite{DeDomenico2014}. The network comprises $N = 3$ layers, representing different routes (underground, overground and DLR) between $n = 369$ train stations.The data also contains  geographical coordinates (latitude and longitude) of train stations. We derive ground truth clusters by performing $k$-means clustering on this geographic data.

In Table \ref{tbl:mltplx} we summarize the important statistics and information of real-world multiplex networks used in our experiments.

\subsubsection{Multiplex biological networks}

We obtained multiplex biological networks of 3 different species, i.e., human biological network (HBN), yeast biological network (YBN) and mouse biological network (MBN)\footnote{The network can be retrieved from: \url{http://morrislab.med.utoronto.ca/~sara/SW/}}, from the study of {\em Mostafavi and Morris} \cite{Mostafavi2010}. The network layers are constructed from the data obtained from different experimental studies and from the publicly available databases. The network layers represent different types of interactions between genes\footnote{genes and their coded proteins are considered as the same type of nodes in networks layers}, including protein interactions, genetic interaction, gene co-expressions, protein localization, disease associations, etc. The number of nodes and layers in each network is summarized in Table \ref{tbl:mltplx}. For each network and its genes, the corresponding GO annotations has also been provided by Mostafavi and Morris \cite{Mostafavi2010}. For details about network statistics and layer representation we refer a reader to {\em Mostafavi and Morris} \cite{Mostafavi2010}.

% {\parindent5mm \textit{Human Biological Networks (HBN)}}: the network data are taken from the study of Didier {\em et al.}\cite{Didier2015}. The network consists of $N = 4$ layers representing different sources of interactions between genes. Namely, protein-protein interaction, mRNA co-expression network, network of pathways and a network of protein complexes. Each network consist of different number of genes and interactions. To construct multiplex network having the same number of nodes in each layer, we take  $n = 1,219$ nodes (genes) that are in intersection of all four networks and for each layer we extract the corresponding interactions between them. The ground truth clustering assignment is unknown, but we use Gene Ontology, a standardized and widely adopted information of gene functions, to validate our clusters.

% {\parindent5mm \textit{Yeast Biological Networks (YBN)}}: the network data are taken from the study of Gligorijevi\'c {\em et al.}\cite{Gligorijevic2014}. The network consist of $N = 4$ layers representing different molecular interactions between $n = $ genes. Namely, protein-protein and genetic interactions, gene co-expression and YeastNet. The modules are validated in the same way as for HBN. 

\begin{table}
\begin{center}
\caption{Real-world multiplex networks used for our comparative study.}
\label{tbl:mltplx}
\begin{tabular}{l|l|c|l|c}
\hline
Net name & $n$ & $N$ & ground truth & Ref.\\
\hline
CiteSeer & 3,312 & 2 &  known ($ k = 3$) & \cite{cora}\\
CoRA & 1,662 & 2 & known ($k = 3$) & \cite{cora}\\ 
MPD & 87 & 3 & known ($k = 6$) & \cite{Dong2012}\\
SND & 71 & 3 & known ($k = 3$) & \cite{CLP}\\
WBN & 279 & 10 & known ($k = 10$) & \cite{White1986}\\
WTN & 183 & 14 & known ($k = 5$) & \cite{De2015}\\
%LTN & 369 & 3 & known ($k = 3$) & \cite{DeDomenico2014}\\
\hline
HBN & 13,251 & 8 & unknown ($k = 100$) & \cite{Mostafavi2010}\\
YBN & 3,904 & 44 & unknown ($k = 100$) & \cite{Mostafavi2010}\\
MBN & 21,603 & 10 & unknown ($k = 100$) & \cite{Mostafavi2010}\\
\hline
\end{tabular}
\end{center}
\end{table}

\subsection{Setup for state-of-the-art methods}

Each of the {\em state-of-the-art} method takes as an input parameter the number of clusters $k$ that needs to be known in advance. Also, some of the methods take as input other types of parameters that needs to be determined. To make the comparison fair, below we briefly explain each of the comparing method and provide the implementation and parameter fitting details that we use in all our experiments (for detailed procedure on parameter fitting, please refer to Section 3 in the online supplementary material):

{\parindent5mm\textit{Baseline, single-layer methods (MM, SNMF, PNMF, SNMTF and SsNMTF)}.} In order to apply them on multiplex network we first merge all the network layers into a single network described by the following adjacency matrix: $\mathbf{A} = \frac{1}{N} \sum_{i=1}^{N} \mathbf{A}^{(i)}$.  

{\parindent5mm\textit{PMM}\cite{Tang2009}} has a single parameter, $\ell$, which represents the number of structural features to be extracted from each network layer. In all our runs, we compare the clustering performance by varying this parameter, but we also noted that the clustering performance does not change significantly when $l \gg k$. 

{\parindent5mm\textit{SNF} \cite{Wang2014}} the method is parameter-free. However, the method prefers  data in the kernel matrix. Thus, we use diffusion kernel matrix representation of binary interaction networks as an input to this method. 

{\parindent5mm\textit{SC-ML}\cite{Dong2014}} has a single regularization parameter, $\alpha$ , that balances the trade-off between two terms in the SC-ML objective function. In all our experiments we choose the value of $\alpha$ that leads to the best clustering performance. 

{\parindent5mm\textit{LMF}\cite{Tang2009lmf}} has a regularization parameter, $\alpha$, that balances the influence of regularization term added to objective function to improve numerical stability and avoid over fitting. We vary $\alpha$ in all our runs, and choose the value of $\alpha$ that leads to the best clustering performance.

{\parindent5mm\textit{GraphFuse}\cite{Papalexakis2013}} has a single parameter, sparsity penalty factor $\lambda$, that is chosen by exhaustive grid search and the value of $\lambda$ that leads to the best clustering performance is chosen.

{\parindent5mm\textit{CGC-NMF}\cite{Cheng2013}} has a set of parameters $\gamma_{ij} \geq 0$  that balance between single-domain and cross-domain clustering objective for each pair of layers $ij$. Given that in all our experiments the relationship between node labels for any pair of layers is {\em one-to-one}, we set $\gamma_{ij} = 1$ (as in \cite{Cheng2013}) for all pairs of layers and throughout all our experiments. 

\subsection{Clustering evaluation measures}

Here we discuss the evaluation measures used in our experiments to evaluate and compare the performance of our proposed methods with the above described {\em state-of-the-art} methods. Given that we test our methods on multiplex network with known and unknown ground truth cluster, we distinguish between two sets of measures:

{\parindent5mm\textit{Known ground truth}.} For multiplex network with known ground truth clustering assignment, we use the following three widely used clustering accuracy measures: {\em Purity}\cite{Zhao2004}, {\em Normalized Mutual Information (NMI)}\cite{Manning2008} and {\em Adjusted Rand Index (ARI)}\cite{Manning2008}. All three measures provide a quantitative way to compare the computed clusters $\Omega = \{\omega_1,\dots, \omega_k\}$  with respect to the ground truth classes: $C = \{c_1,\dots, c_k\}$. {\em Purity} represents percentage of the total number of nodes classified correctly, and it is defined as \cite{Zhao2004}:

\begin{equation*}
Purity(\Omega,C) = \frac{1}{n}\sum_{k} \max_{j} |\omega_k \cap c_j|
\end{equation*} 

\noindent where $n$ is the total number of nodes, and $|\omega_k \cap c_j|$ represents the number of nodes in the intersection of $\omega_k$ and $c_j$. To trade-off the quality of the clustering against the number of clusters we use {\em NMI}. NMI is defined as \cite{Manning2008}:

\begin{equation*}
NMI(\Omega, C) = \frac{I(\Omega; C)}{|H(\Omega)+ H(C)|/2}
\end{equation*} 

\noindent where $I$ is the mutual information between node clusters $\Omega$ and classes $C$, while $H(\Omega)$ and $H(C)$ represent the entropy of clusters and classes respectively. Finally, {\em Rand Index} represents percentage of true positive ($TP$) and true negative ($TN$) decisions assigns that are correct (i.e., accuracy). It is defined as:

\begin{equation*}
RI(\Omega,C) = \frac{TP + TN}{TP + FP + FN + TN}
\end{equation*}  

\noindent where, $FP$ and $FN$ represent false positive and false negative decisions respectively. {\em ARI} is defined to be scaled in range $[0,1]$ \cite{Manning2008}. All three measures are in the range $[0,1]$, and the higher their value, the better clustering quality is.

{\parindent5mm\textit{Unknown ground truth}.} For biological networks, the ground truth clusters are unknown and evaluating the clustering results becomes more challenging. In order to evaluate the functional modules identified by our methods, we use Gene Ontology (GO) \cite{Ashburner2000}, a commonly used gene annotation database. GO represents a systematic classification of all known protein functions organized as well-defined terms (also known as GO terms) divided into three main categories, namely Molecular Function (MF), Biological Process (BP) and Cellular Component (CC) \cite{Ashburner2000}. GO terms (i.e., annotations), representing gene functions, are hierarchically structured where low-level (general) terms annotate more proteins than high-level (specific) terms. Thus, in our analysis we aim to evaluate our clusters with high-level (specific) GO terms annotating not more than 100 genes. Additionally, we remove GO terms annotating 2 or less proteins. Thus, for each gene in a network, we create a list of its corresponding GO term annotations. We then analyze the consistency of each cluster, $i$, obtain by our method, by computing the {\em redundancy} \cite{Pereira2004}, $R_i$ as follows:

\begin{equation*}
R_{i} = 1 - \frac{\Big (-\sum\limits_{l=1}^{N_{GO}} p_l \log_2 p_l \Big )}{\log_2 N_{GO}}
\end{equation*}
\noindent where, $N_{GO}$ represents the total number of GO terms considered and $p_l$ represents the relative frequency of GO term $l$ in cluster $i$. Redundancy is based on normalized Shannon's entropy and its values range between 0 and 1. For clusters in which the majority of genes share the same GO terms (annotations) redundancy is close to 1, whereas for clusters in which the majority of genes have disparate GO terms the redundancy is close to 0. When comparing clustering results obtained by different methods, we use the value of redundancy averaged over all clusters. Furthermore, the redundancy is a very suitable measure for clustering performance comparisons because its value does not depend on the number of clusters and unlike others evaluation measures for biological network clustering \cite{Shih2012}, it is parameter-free.

\begin{figure}
\centering
\includegraphics[scale=0.4]{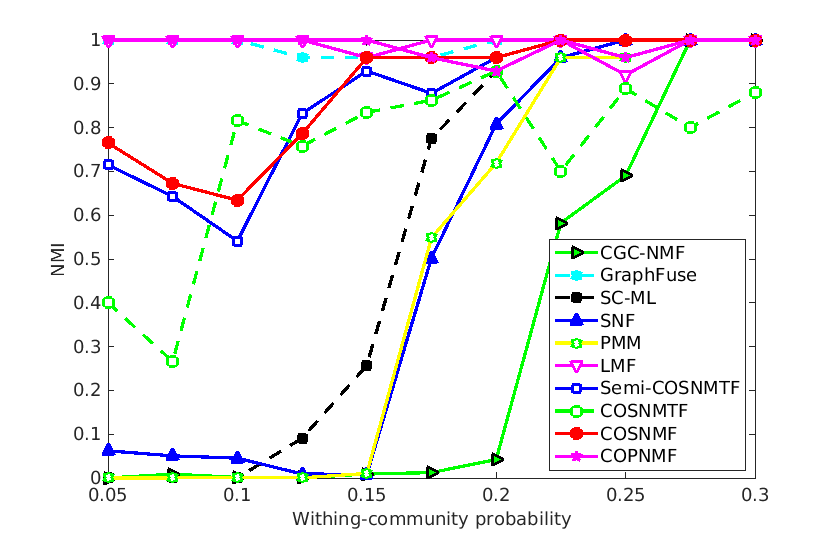}
\caption{The clustering performance of our proposed and other methods on 11 different {\em SYNTH-C} multiplex networks measured by NMI. On $x$-axis we present within-community probability, representing the density of connections of communities in the two complementary layers.}
\label{fig:synth-c}
\end{figure}

% The percentage of genes in a biological network covered by functionally consistent clusters, as well as clusters functional diversity play important roles in many biological applications. Thus, in addition to redundancy, we also compute {\em network coverage} and cluster {\em functional diversity} \cite{Han2004}. The network coverage, $Cov$, represents the fraction of nodes covered by functionally consistent clusters, i.e., $Cov = \frac{1}{n} \sum\limits_{i=1}^k |c_i (R_i > 0)|$, where $|c_i (R_i > 0)|$ represents the number of nodes in a functionally consistent cluster $i$; we consider a cluster $c_i$ to be functionally consistent if $R_i > 0$. The functional diversity of each cluster $i$, $D_i$, is computed as a number of unique GO terms of a cluster divided by the number of nodes in the cluster.  

\section{Results and Discussion}\label{sec:results}

\subsection{Clustering evaluation on artificial multiplex networks}

The ability of our proposed methods to extract clusters from complementary layers, represented by {\em SYNTH-C} networks, is shown in Figure \ref{fig:synth-c}. The performance of our methods is compared with other methods and it is measured by NMI. By decreasing the within-community probability of complementary clusters in both layer, i.e., by decreasing the density of connections withing communities and thus making communities harder to detect, we see a drastic decrease of performance in many methods, including SC-ML, PMM, SNF and CGC-NMF (Fig. \ref{fig:synth-c}). Furthermore, below some value of within-community probability, i.e., $< 0.1$, the performance of these methods is equal or close to zero. Unlike them, our proposed methods, particularly CSNMF, CSsNMTF and CPNMF show significantly better performance. Specifically, CPNMF demonstrates constant performance for all values of withing-community probability. The similar results can also be observed for GraphFuse and LMF. Given that, we can observe that CPNMF method is mostly successful in utilizing complementary information contained in all layers  and achieving the highest clustering results.      

\begin{figure}[!b]
\centering
\includegraphics[scale=0.4]{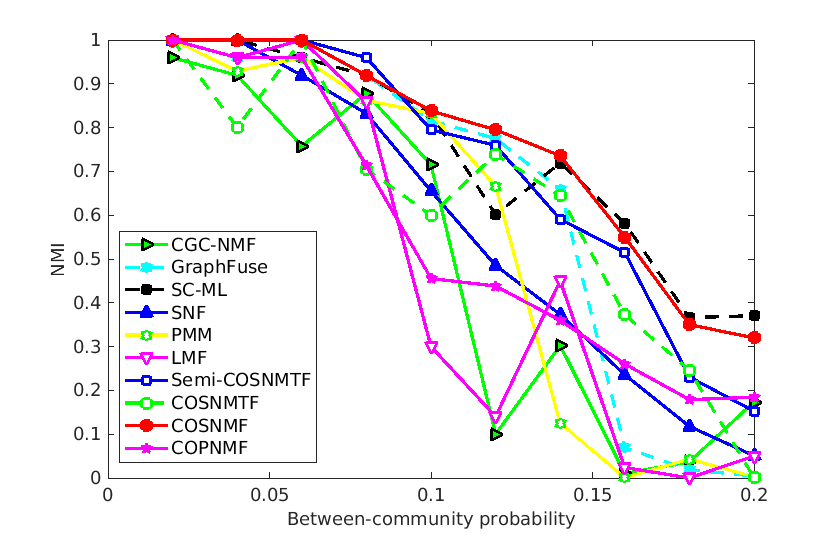}
\caption{The clustering performance of our proposed and other methods on 10 different {\em SYNTH-N} multiplex networks measured by NMI. On $x$-axis we present between-community probability, representing the noise level between communities in the first layer.}
\label{fig:synth-n}
\end{figure}

In terms of noise, incorporated into {\em SYNTH-N} networks, the ranking between the methods in terms of clustering performance is different. By increasing the between-community probability of the first layer, and thus introducing more noise between communities, the clustering performance of all methods decreases (Fig. \ref{fig:synth-n}). Our proposed methods, CSNMF, CSNMTF and CSsNMTF, along with SC-ML demonstrate the best performance across different values of within-community probability, which makes them more robust to noise than other methods. On the other hand, other methods methods are characterized with significantly lower clustering performance. Surprisingly, CPNMF method, which is giving the best performance for complementary layers, perform significantly worse on nosy networks than other methods.   

\subsection{Clustering evaluation on real-world multiplex networks}

\begin{table*}[t]
\begin{center}
\caption{Clustering accuracy measures for methods (from left to right): MM, SNMF, PNMF, SNMTF, SsNMTF, PMM, SNF, SC-ML, LMF, GraphFuse, CGC-NMF, CSNMF, CPNMF, CSNMTF, CSsNMTF applied on real-world multi-layer networks (from top to bottom): {\em CiteSeer}, {\em CoRA}, {\em MPD}, {\em SND}, {\em WBN}, {\em WTN}, {\em HBN}, {\em YBN} and {\em MBN}}.
\label{tbl:real}
\resizebox{\linewidth}{!}{%
\begin{tabular}{l|c|c|c|c|c|c|c|c|c|c|c|c|c|c|c|c} 
\hline
    & & MM & SNMF & PNMF & SNMTF & SsNMTF &  PMM  & SNF & SC-ML & LMF & GF &  CGC & CSNMF & CPNMF & CSNMTF & CSsNMTF\\
\hline
%\rowcolor{lightgray}
\parbox[b]{2mm}{\multirow{3}{*}{\rotatebox[origin=c]{90}{{\scriptsize CiteSeer}}}} 
& Purity & 0.500 & 0.407 & 0.405 & 0.377 & 0.371 & 0.302 & 0.214 &    0.419 & 0.235 & 0.512 & 0.212 & 0.501 & {\bf 0.519} & 0.416 &  0.404\\
%\rowcolor{lightgray}
& NMI & 0.187 & 0.222 & 0.221 & 0.170 & 0.164 & 0.145 & 0.023 & 0.191 & 0.013 & 0.211 & 0.013 & {\bf 0.237} & 0.216 & 0.172 & 0.195\\
%\rowcolor{lightgray}
& ARI & 0.152 & 0.059 & 0.057 & 0.042 & 0.038 & 0.008 & 0.001 & 0.169 & 0.005 &  0.201 & 0.001 & {\bf 0.207} & 0.185 & 0.093 & 0.089\\
\hline\hline
\parbox[b]{2mm}{\multirow{3}{*}{\rotatebox[origin=c]{90}{{\scriptsize CoRA}}}} 
& Purity & 0.706 & 0.669 & 0.660 & 0.669 & 0.669 & 0.496 & 0.733 & 0.787 & 0.492 & 0.642 & 0.678 & {\bf 0.802} & 0.790 & 0.683 & 0.684\\   
& NMI & 0.340 &  0.385 & 0.353 & 0.382 & 0.382 & 0.085 & 0.449 & 0.480 & 0.002 &    0.201 & 0.389 &  {\bf 0.514} & 0.480 & 0.346 & 0.390\\
& ARI & 0.257 & 0.280 & 0.247 & 0.277 & 0.277 & 0.030 & 0.470 & 0.485 & 0.001 &    0.209 & 0.296 & {\bf 0.491} & 0.470 & 0.279 & 0.288\\ 
\hline\hline
%\rowcolor{lightgray}
\parbox[b]{2mm}{\multirow{3}{*}{\rotatebox[origin=c]{90}{{\scriptsize MPD}}}} 
& Purity &  0.563 & 0.678 & 0.666 & 0.620 & 0.678 & 0.689 & 0.620 & 0.701 & 0.471 &  0.689 & 0.678 & 0.701 & 0.655 & 0.655 & {\bf 0.724} \\
%\rowcolor{lightgray}
& NMI & 0.313 & 0.471 & 0.466 & 0.384 & 0.473 & 0.533 & 0.395 & 0.495 & 0.191 & {\bf 0.565} &  0.457 & 0.504 & 0.451 & 0.458 &  0.521 \\
%\rowcolor{lightgray}
& ARI & 0.147 & 0.268 & 0.259 & 0.228 & 0.272 & 0.383 & 0.280 & 0.379 & 0.029 &    0.411 & 0.357 & 0.394 & 0.368 & 0.346 & {\bf 0.422}\\
\hline\hline
%\rowcolor{lightgray}
\parbox[b]{2mm}{\multirow{3}{*}{\rotatebox[origin=c]{90}{{\scriptsize SND(o)}}}} 
& Purity & 0.929 & {\bf 0.943} & {\bf 0.943}  & 0.676 & {\bf 0.943} & {\bf 0.943} & {\bf 0.943} & {\bf 0.943} & 0.788 & {\bf 0.943} & {\bf 0.943} &  {\bf 0.943} & {\bf 0.943} & {\bf 0.943} & {\bf 0.943} \\
%\rowcolor{lightgray}
& NMI &  0.618 & 0.681 &  0.681 & 0.133 & 0.681 & 0.675 & 0.689 & 0.681 & 0.303  & 0.675 & 0.673 & 0.681 & 0.685 & {\bf 0.773} & 0.678\\
%\rowcolor{lightgray}
& ARI &  0.460 & 0.493 & 0.493 & 0.021 &  0.493 & 0.477 & 0.515 & 0.493 & 0.239 &  0.477 &  0.472 & 0.493 &  0.503 & {\bf 0.811} & 0.484\\ 
\hline\hline
%\rowcolor{lightgray}
\parbox[b]{2mm}{\multirow{3}{*}{\rotatebox[origin=c]{90}{{\scriptsize SND (s)}}}} 
& Purity & 0.619 & 0.577 & 0.633 & 0.634 & 0.619 & 0.591 & 0.633 & 0.591 & 0.633 &    0.619 & 0.662 & 0.633 & 0.633 & {\bf 0.747} &  0.605\\ 
%\rowcolor{lightgray}
& NMI & 0.038 & 0.025 & 0.052 & 0.055 & 0.041 & 0.037 & 0.057 & 0.030 & 0.053 &    0.045 & 0.0781 & 0.053 & 0.053 & {\bf 0.276} & 0.034\\
%\rowcolor{lightgray}
& ARI & 0.024 & 0.012 & 0.058 & 0.058 & 0.043 & 0.022 &  0.058 & 0.021 & 0.059 &    0.044 & 0.092 & 0.059 & 0.059 &  {\bf 0.234} & 0.031\\ 
\hline\hline
%\rowcolor{lightgray}
\parbox[b]{2mm}{\multirow{3}{*}{\rotatebox[origin=c]{90}{{\scriptsize WBN}}}} 
& Purity & 0.473 & 0.512 & 0.501 & 0.476 & 0.523 & 0.473 & 0.534 & 0.272 & 0.283 &  0.509 & 0.516 & {\bf 0.577} & 0.537 & 0.548  & 0.530\\
%\rowcolor{lightgray}
& NMI & 0.333 & 0.382 & 0.400 & 0.327 & 0.363 & 0.373 & 0.425 & 0.079 & 0.098 & 0.426 & 0.370 &  {\bf 0.463} & 0.432  & 0.404 &  0.424\\
%\rowcolor{lightgray}
& ARI &  0.199 & 0.226 & 0.213 & 0.112 & 0.180 & 0.130 & 0.211 & 0.001 & 0.009 & 0.216 &    0.211 & {\bf 0.291} & 0.233 & 0.237 & 0.225 \\
\hline\hline
%\rowcolor{lightgray}
\parbox[b]{2mm}{\multirow{3}{*}{\rotatebox[origin=c]{90}{{\scriptsize WTN}}}} 
& Purity & 0.506 & 0.475 &  0.464 & 0.388 & 0.284 & 0.388 & 0.289 & 0.497 & 0.453 & 0.415 & 0.278 &  {\bf 0.579} &  0.420 & 0.371 & 0.420 \\
%\rowcolor{lightgray}
& NMI & 0.231 & 0.269 & 0.242 & 0.176 & 0.077 & 0.205 & 0.073 & 0.226 & 0.191 & 0.176 & 0.072 &   {\bf 0.322} & 0.172 & 0.154 & 0.155 \\
%\rowcolor{lightgray}
& ARI & 0.080 & 0.114 & 0.114 & 0.073 & 0.001 & 0.039 & 0.005 & 0.133 & 0.094 & 0.107 & 0.002 &    {\bf 0.160} & 0.094 & 0.035 & 0.088\\
\hline\hline
{\scriptsize HBN} &  &  0.180 & 0.325 & 0.350 & 0.322 & 0.326 & 0.351 & 0.141 & 0.266 & 0.045 & 0.203 & 0.320 & 0.341 & {\bf 0.364} & 0.339 & 0.342\\
{\scriptsize YBN} & $R_{avg}$ & 0.027 & 0.374 & 0.336 & 0.372 & 0.358 &    0.343 & 0.163 & 0.257 & 0.100 & 0.114 & 0.311 & {\bf 0.383} & 0.342 & {\bf 0.383} & 0.381\\
{\scriptsize MBN} &  & 0.015 &  0.416 & 0.387 & {\bf 0.462} & 0.441 & 0.355 & 0.211 & 0.320 & 0.180 & 0.298 & 0.328 & 0.422 & 0.401 & 0.416 & 0.433\\
\hline\hline
\end{tabular}}
\end{center}
\end{table*}

% \begin{figure}
% \centering
% \includegraphics[scale=0.5]{figs/cov_and_div}
% \caption{(A) Network coverage and (B) average functional diversity of clusters for HBN, YBN and MBN computed by our four methods, {\em CSNMF}, {\em CPNMF}, {\em CSNMTF} and {\em CSsNMTF}.}
% \label{fig:covdiv}
% \end{figure}

Table \ref{tbl:real} presents the Purity, NMI and ARI of our four proposed collective factorization methods, along with five different baseline methods and six different widely used {\em state-of-the-art} methods on six different real-world networks. The first important observation is that all four proposed collective NMF methods (CSNMF, CPNMF, CSNMTF and CSsemi-NMTF) perform better than their corresponding baseline methods (SNMF, PNMF, SNMTF and Ssemi-NMTF) on all real-world multiplex networks. Thus, the strategy of merging layers into a single layer always leads to underperformance. Moreover, single-layer modularity maximization (MM) algorithm is outperformed by baseline, single-layer NMF methods in almost all real-world networks, except for WTN networks where MM significantly outperforms baseline NMF methods, and SND(o) where MM performs better than SNMF, SNMTF and Ssemi-NMTF, but not better than PNMF. In comparison to the {\em state-of-the-art} methods (PMM, SNF, SC-ML, LMF, GraphFuse and CGC-NMF), at least one of our proposed methods outperforms them all (in terms of either Purity, NMI or ARI or all three measures) in all real-world multiplex network. Moreover, for example, on MPD network, both CSNMF and CSsemi-NMTF perform better than all other methods, with CSemi-NMTF being the best in terms of Purity and NMI; on SND(s) network, CSNMF, CSNMTF and CSsemi-NMTF perform better than all other methods, with CSNMTF performing the best in terms of all three measures; on WBN network, both CSNMF and CSNMTF perform better than all other methods, with CSNMTF being the best in terms of Purity and ARI, and CSNMF being the best in terms of NMI; on WTN network, CSNMF, CSNMTF and CSsemi-NMTF perform better than other methods, with CSNMF being the best in terms of all three measures. 

\subsubsection{Clustering evaluation on multiplex biological networks}

In table \ref{tbl:real}, we also present the average redundancy ($R_{avg}$) obtained by clustering multiplex biological networks with our four methods, as well as with {\em state-of-the-art} and baseline methods. The results, again, indicate the superior performance of our methods over the {\em state-of-the-art} and baseline methods, except in the case of MBN, where SNMTF, applied on merged network layers, yields the highest redundancy. 
% When comparing our methods within themselves in terms of network coverage, we observe that {\em CPNMF} often leads to lower coverage than other three methods (see Fig. \ref{fig:covdiv} (A)); whereas in terms of functional diversity averaged over clusters, all four methods produce comparable results, with {\em CSsNMTF} producing slightly better results than the other three methods (see Fig. \ref{fig:covdiv} (B)).

\begin{figure}
\centering
\includegraphics[scale=0.5]{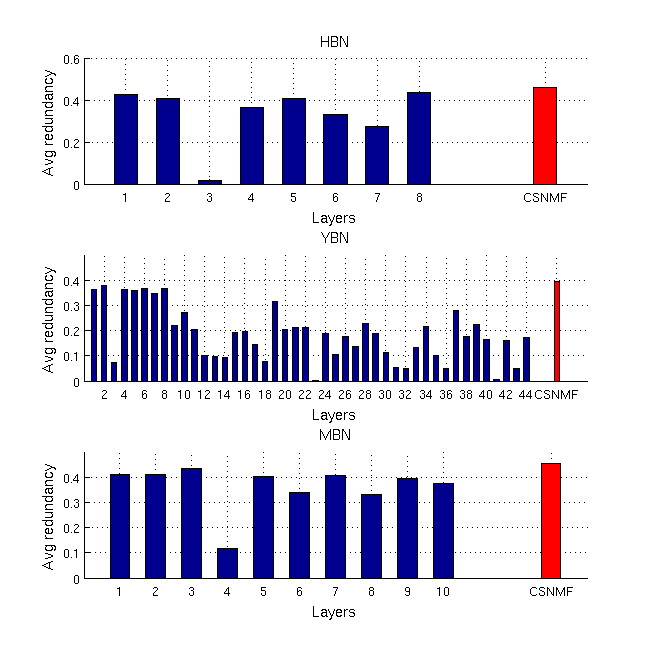}
\caption{Average redundancy of each individual network layer (in blue), computed by {\em SNMF}, and of their fused representation (in red), computed by {\em CSNMF}, for networks: (A) HBN, (B) YBN and (C) MBN. Method parameters are: $k = 300$ and $\alpha = 0.01$.}
\label{fig:redund}
\end{figure}

Furthermore, we compare the functional consistency of clusters obtain by collective integration of all network layers with the functional consistency of clusters obtained from each individual network layer. The results for all three biological networks, obtained by applying SNMF on each individual network layer and CSNMF on all network layers together, are depicted in Fig. \ref{fig:redund}. We observe that each individual network layer has biologically consistent clusters. However, the highest level of biological consistency, measured by the average redundancy, is achieved when all the layers are fused together (red bars in Fig.\ref{fig:redund}). 

\section{Conclusion}\label{sec:conc}

In this paper, we address the problem of composite community detection in multiplex networks by proposing NF-CCE, a general model consisting of four novel methods, {\em CSNMF}, {\em CPNMF}, {\em CSNMTF} and {\em CSsemi-NMTF}, based on four non-negative matrix factorization techniques. Each of the proposed method works in a similar way: in the first step, it decomposes adjacency matrices representing network layers into low-dimensional, non-negative feature matrices; then, in the second step, it fuses the feature matrices of layers into a consensus non-negative, low-dimensional feature matrix common to all network layers, from which the composite clusters are extracted. The second step is done by collective matrix factorization that maximizes the shared information between network layers by optimizing the distance between each of the non-negative feature matrices representing layers and the consensus feature matrix. 

The ability of our methods to integrate complementary as well as noisy network layers more efficiently than the {\em state-of-the-art} methods has been demonstrated on artificially generated multiplex networks. In terms of clustering accuracy, we demonstrate the superior performance of our proposed methods over the baseline and {\em state-of-the-art} methods on nine real-world networks. We show that simple averaging of adjacency matrices representing network layers (i.e., merging network layers into a single network representation), the strategy that is usually practiced in the literature, leads to the worst clustering performance. Moreover, our experiments indicate that widely-used modularity maximization methods are significantly outperformed by NMF-based methods. 

NF-CCE can be applied on multiplex networks from different domains, ranging from  social, phone communication and bibliographic networks to biological, economic and brain networks, demonstrating the diverse applicability of our methods.  

\ifCLASSOPTIONcaptionsoff
  \newpage
\fi



%\bibliographystyle{IEEEtran}
\bibliographystyle{IEEEtran}
\bibliography{IEEEabrv,biblio.bib}

\end{document}